\documentclass[preprint,epsfig]{aastex}
\usepackage{times}
\usepackage{graphicx}        
\begin{document}

\title{Narrowing of the neutrino light curve in the OPERA experiment}
\author{Maurice H.P.M. van Putten}
\affil{Korea Institute for Advanced Study, Dungdaemun-gu, Seoul 130-722, Korea,  \email{mvputten@kias.re.kr}  }

\begin{abstract}
In the OPERA experiment, the time of arrival of the neutrino light curve at LNGS is defined by a best-fit to the generating proton PDF at CERN. By a two parameter matched filtering procedure, we determine that the OPERA neutrino light curve is narrower by about 0.4-0.5\% relative to the proton PDF. The results indicate as yet undetermined physics in the creation of neutrinos in CERN and a reduced significance of the OPERA detection to less than 4 $\sigma.$
\end{abstract}

\maketitle

\section{Introduction}

The OPERA experiment \citep{ada11} recently reported an advance arrival time of muon neutrinos in traversing a distance $D=$730 km between CERN and the detector at LNGS. The OPERA collaboration reports a discovery with 6 $\sigma$ significance of an excess velocity of $(2.48\pm 0.28(stat)\pm 0.30 (sys))\times 10^{-5}$ relative to the velocity of light. Some support for this result comes from a similar outcome of the MINOS experiment \citep{ada07} over essentially the same baseline distance and operating at similar energies. If true, these results should be considered in contrast to a null-result for the $<20$ MeV neutrino detections from SN1987A \citep{bur87}.

MINOS \citep{ada07} performed an early attempt with relatively few neutrinos to measure the true neutrino time-of-flight (TOF) by measuring arrival times at the locations of two neutrino detectors. This method is insensitive to the nature of neutrino creation. OPERA, instead, measures a TOF indirectly on the basis of a template, defined by the proton PDF. The arrival time of the neutrino bursts at LNGS is defined by a one-parameter best-fit between the template and the neutrino light curve by application of matched filtering.

\begin{center}
\begin{figure}[ht]
\center{\includegraphics[angle=00,scale=.3]{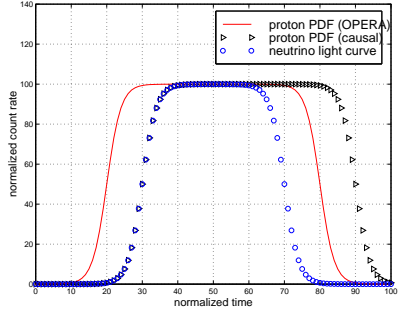}}
\caption{Schematic overview of extracting the time of arrival from a best-fit of a template $(red)$ to the neutrino light curve ($blue~circles$), where the template is defined by the proton PDF from CERN and the latter obtains from the detector at LNGS. Using time shifts only, OPERA uses a centered matching filtering that, in case of discrepant widths in the neutrino burst and the template, produces an apparent advanced arrival time. Taking into account causality, an additional shift is required ($triangles$), that aligns the front of the template to the front of the neutrino light curve. The additional shift can be obtained by calculating the relative width of the neutrino light curve in a two-parameter matched filtering procedure.}
\label{FIG_0}
\end{figure}
\end{center}

Extracting the true TOF to evaluate the possibility of signals traveling faster than the velocity of light requires taking into account causality, here the time of arrival of the front leading the neutrino burst of 10.5 $\mu$s. The OPERA defines the time of arrival of the neutrino burst by a best-fit position of the proton PDF to the neutrino light curve on the basis of time shifts. As illustrated in Fig. \ref{FIG_0}, this procedure is sensitive to the width of the neutrino burst relative to the width of the proton PDF, that may be different due to uncertainties in the process of neutrino creation. In particular, a neutrino light curve that is narrower by 1.2\% relative to the proton PDF produces an apparent 60 ns advance arrival time, as inferred from an inadvertently advanced position by 0.6\% of the front of the proton PDF produced in the OPERA best-fit analysis relative to the front of the neutrino light curve. 

Here, we present a refined analysis of the OPERA data-analysis by a two-parameter matched filtering procedure, that identifies and corrects for discrepant widths of the neutrino light curve and to the proton PDF. We apply matched filtering comprising both time shifts and stretching, developed in the analysis of light curves of long duration gamma-ray bursts \citep{van09}.
\begin{center}
\begin{figure}[ht]
\center{\includegraphics[angle=00,scale=.3]{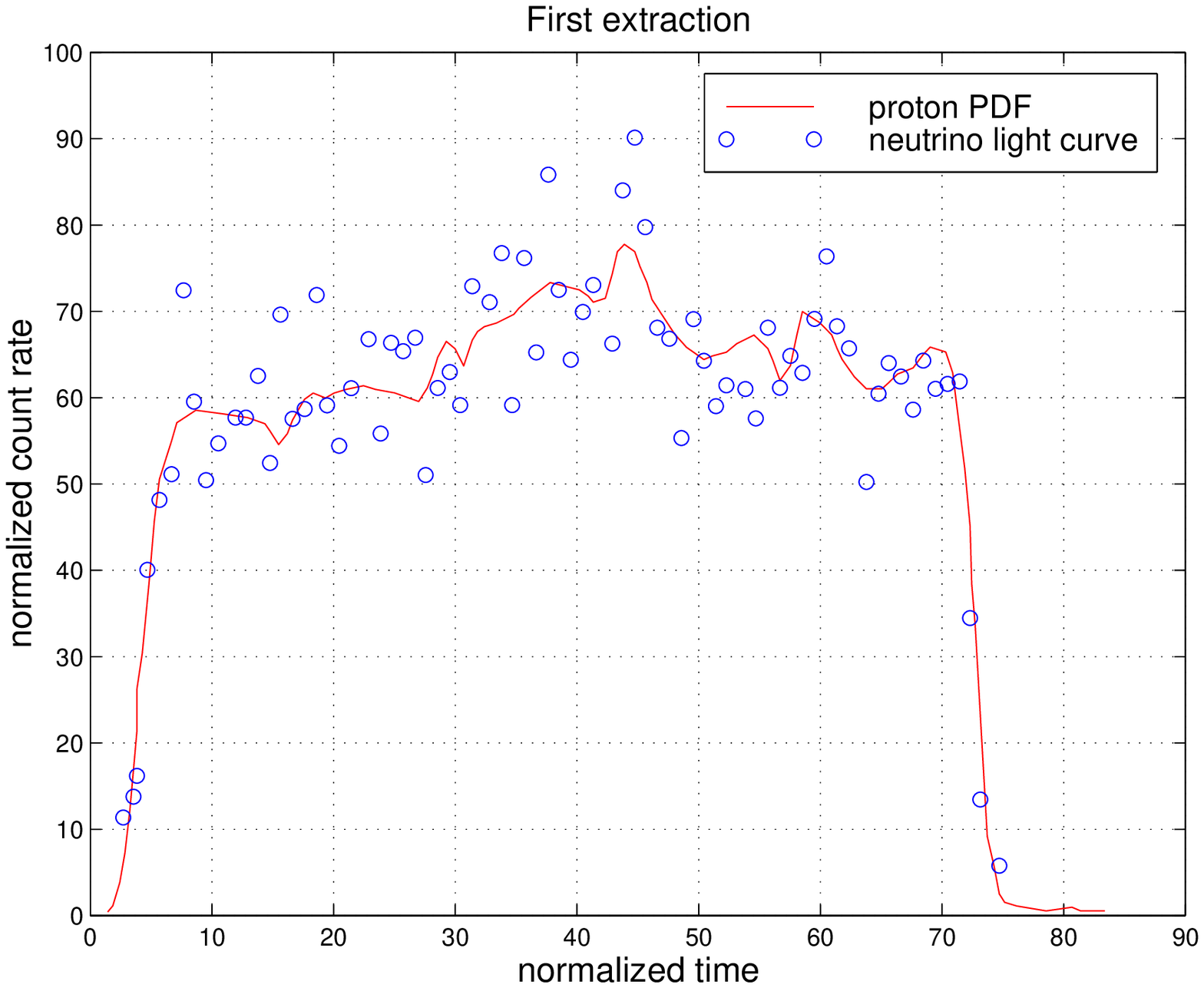}\includegraphics[angle=00,scale=.3]{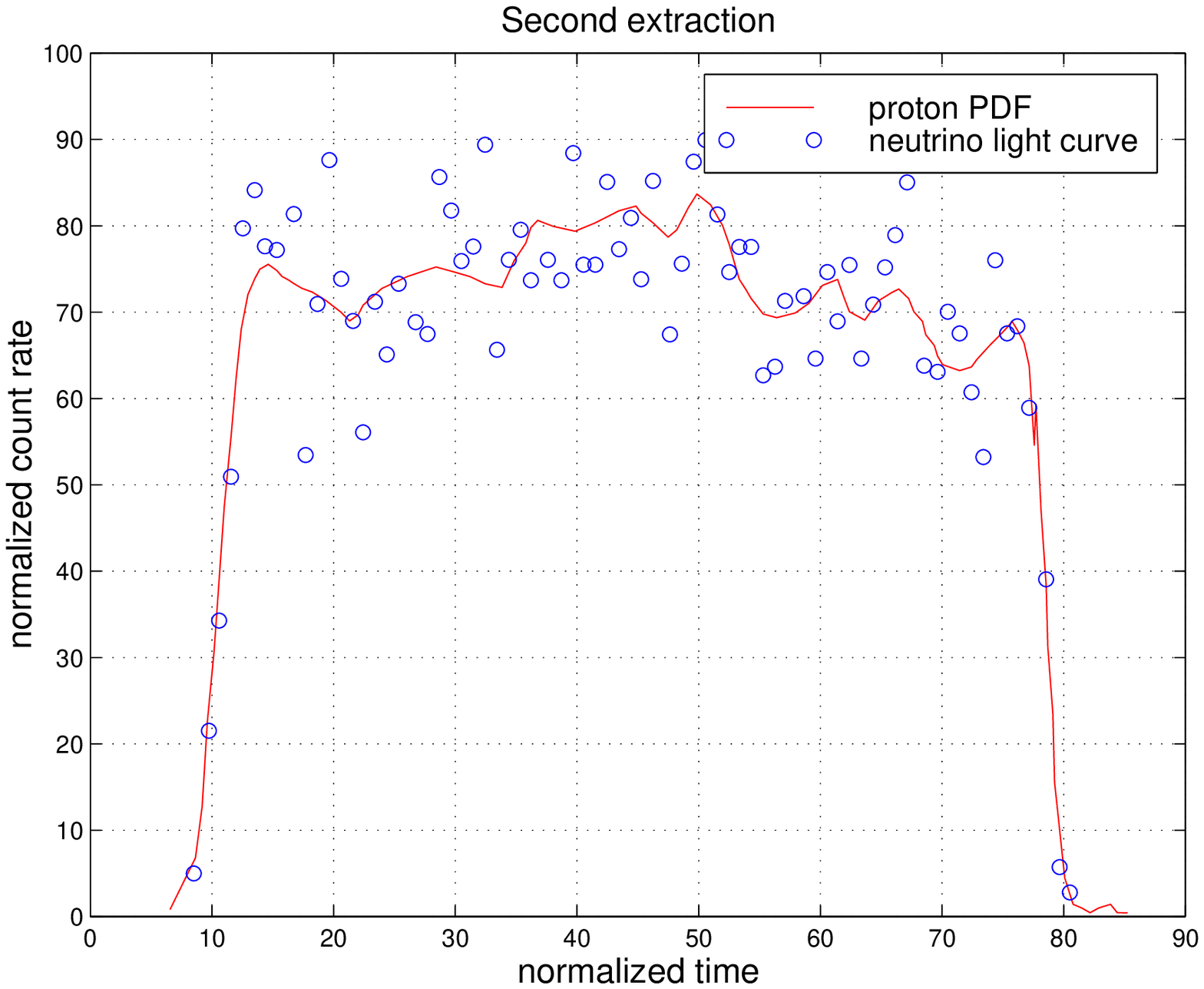}}
\caption{Shown are the First ($left$) and Second ($right$) extractions of the proton PDF ($red$) and the neutrino light curves $(blue~ circles)$ in the OPERA experiment. (Extracted from \cite{ada11}.)}
\label{FIG_1}
\end{figure}
\end{center}

The OPERA experiment measurement of the light curve in neutrinos at LNGS in relation to a proton PDF of about 10.5 $\mu$s duration in their creation at CERN represents the accumulation of 16111 events gathered in 2009-2011, shown in Fig. \ref{FIG_1} following \cite{ada11}. 

\begin{center}
\begin{figure}[ht]
\center{\includegraphics[height=60mm,width=60mm]{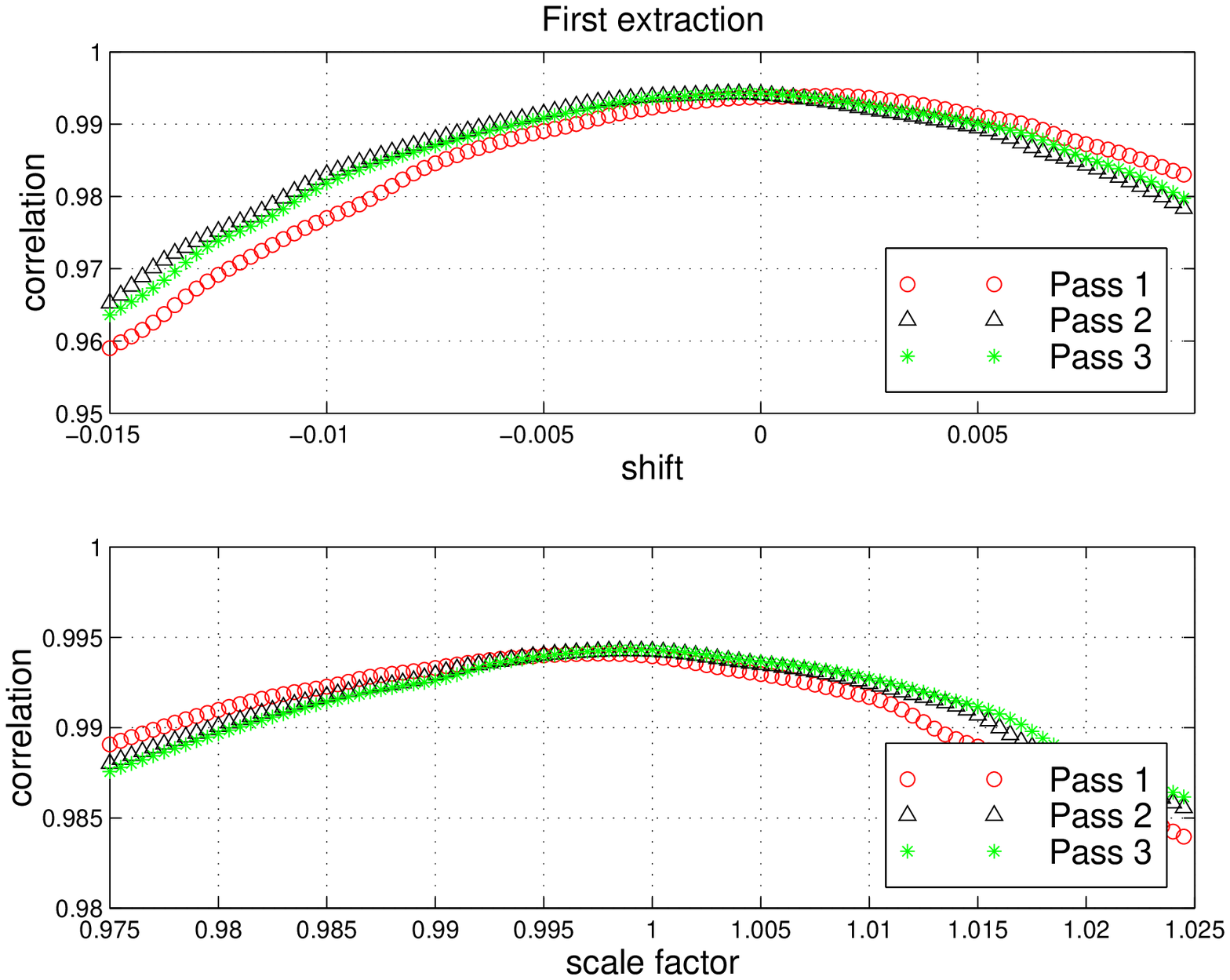}\includegraphics[height=60mm,width=60mm]{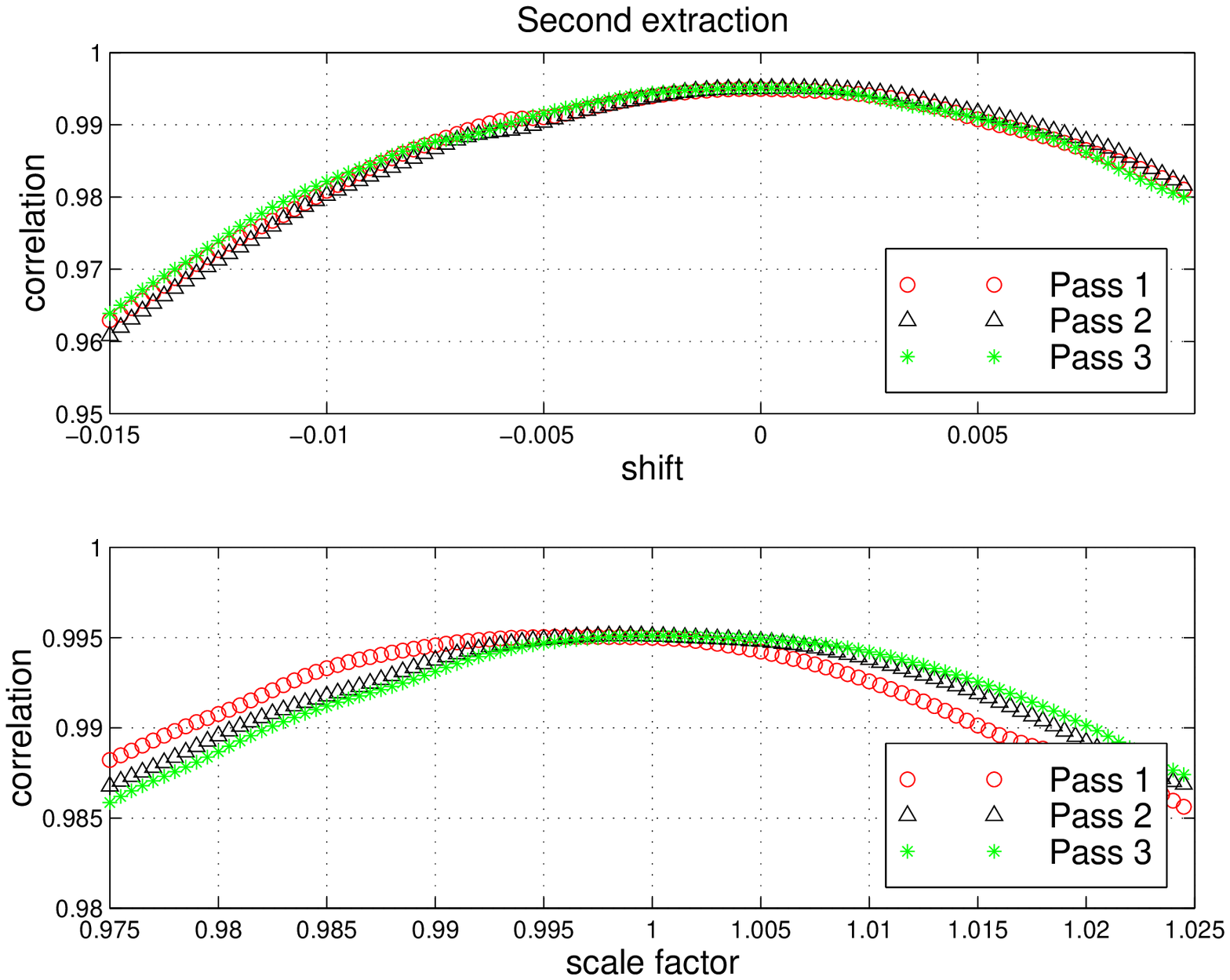}}
\caption{Shown is the two-parameter matched filtering of the proton PDF as a template to the neutrino light curve in the OPERA experiment. The match by shift, $\delta t$, and scale factor, $f_s$, of the template is optimal for $(\delta t,f_s)$=(0.001,0.996) and (0.0005, 0.995) for the first and second extractions, respectively, following three iterations (Pass~1-3). The results for $f_s$ show that the neutrino light curve is about 0.4\% to 0.5\% narrower than the proton PDF. }
\label{FIG_2}
\end{figure}
\end{center}

To determine the width of the neutrino light curve relative to proton PDF, we first normalize the data in durations and count rates and consider the normalized proton PDF as the template in a matched filtering procedure. We search for an optimal fit to the neutrino light curve, by transforming the template by shifts $\delta t$ in time and scaling by $f_s$ of the time axis, and search for the optimal match of the transformed template to the normalized neutrino light curve data. The results are obtained following three iterations shown in Fig. \ref{FIG_2}. 

The OPERA experiment considers First and Second extractions, defined by two successive neutrino bursts separated by 50 ms. The results of Fig. \ref{FIG_2} shows that the width of the normalized neutrino light curve relative to the normalized proton PDF is 0.4\% and 0.5\% in the First and Second extractions. The average of 0.45\% introduces a fractional correction of
$0.45/1.2=0.375$ to the OPERA significance of 6 $\sigma$, reducing it to 3.75 $\sigma$. 

The above shows that the OPERA experiment, in contradistinction to the MINOS experiment, is sensitive to the nature of creation of the neutrinos that may, for as yet undetermined reasons, produce a neutrino light curve that is narrower than the proton PDF. A two-parameter matched filtering analysis of the First and Second extractions shows that this is indeed the case. It introduces a  correction to the one-parameter OPERA data-analysis, reducing the overall significance to less than 4$\sigma.$ 

Our results indicate the need for enhanced data-sets by continuation of MINOS and OPERA experiments and analysis by two or more parameter matched filtering methods. Conceivably, future results point to novel physics in neutrino creation and propagation. 

{\bf Acknowledgments.} The author thanks stimulating discussions with Henry Tye and Eoin O. Colgain.

\end{document}